\documentclass[a4paper]{revtex4}
\usepackage{graphicx}

\begin{document}
\title{Observables have no value:
\\ a no-go theorem for position and momentum observables}
\author{Alberto C. de la Torre }
\email{delatorre@mdp.edu.ar}
 \affiliation{Departamento de F\'{\i}sica,
 Universidad Nacional de Mar del Plata\\
 Funes 3350, 7600 Mar del Plata, Argentina\\CONICET}
\begin{abstract}
The Bell-Kochen-Specker contradiction is presented using
continuous observables in infinite dimensional Hilbert space. It
is shown that the assumption of the \emph{existence} of putative
values for position and momentum observables for one single
particle is incompatible with quantum mechanics.
\end{abstract}
\maketitle
\section{INTRODUCTION}
One of the central questions in the interpretation of quantum
mechanics from a realist perspective is whether the
indeterminacies or uncertainties in quantum mechanics are of an
\emph{ontological} or \emph{gnoseological} character. They are
gnoseological if the observables of the system possess exact
values that quantum mechanics is unable to predict and can only
provide probability distributions for them. In this case the
uncertainty is in our knowledge of the system and not in the
system itself and the development of a deterministic theory with
hidden variables that are averaged to produce the same
statistical predictions of quantum mechanics is wished. The
indeterminacies are ontological if the observables do not assume
exact values but instead they are diffuse by nature and the
indeterminacies are in nature and not in our knowledge. In this
case quantum mechanics can be considered to be a complete theory
and not the statistical average of a better theory. The
Einstein-Podolsky-Rosen argument\cite{epr} was originally
designed in order to prove that quantum mechanics is not
complete, although later developments favour an interpretation of
the argument where the values of the ``elements of physical
reality'' are nonlocality determined (a special case of
contextuality). Indeed, the experimental violations\cite{exp} of
Bell's inequality\cite{bell} have established such nonlocal
effects in the valuation of observables.

The existence of definite context independent values for the
observables was shown by Bell\cite{bell2} and by Kochen and
Specker\cite{koch} to be in conflict with quantum mechanics on
logical grounds, that is, in conflict with the geometrical
structure of the Hilbert space, more than with the postulates of
quantum mechanics. The Kochen Specker theorem is a complicated
argument requiring 117 vectors in a three dimensional Hilbert
space. A simpler proof was produced by Peres\cite{per} with only
33 vectors and Penrose\cite{pen} found a beautiful geometrical
representation for these vectors. With another goal, not trying
to minimize the number of directions, a proof of the theorem was
given\cite{gill} involving continuous sets of directions.
Analysing spin observables for systems of two and three
particles, Mermin\cite{mer} presented physical examples of the
Bell-Kochen-Specker contradiction.

The original Einstein-Podolsky-Rosen argument involves
observables of position and momentum and D. Bohm\cite{Bohm}
presented the same argument but in terms of spin observables.
Since then, spin observables were preferred for
Einstein-Podolsky-Rosen and Bell-Kochen-Specker type of
arguments. This preference is not only because these involve a
finite dimensional Hilbert space with simpler mathematics, but
mainly because spin observables are more adequate for real
experimental tests. However, spin is an essentially quantum
mechanic observable with almost no correlate in classical
mechanics and therefore it is interesting to exhibit these
arguments also for position and momentum observables in order to
emphasize the drastic differences between classical and quantum
mechanics. With these observables, a ``frame function'' was
constructed and a rigourous proof of the Bell-Kochen-Specker
theorem was provided\cite{zimba}. Furthermore, illustrations of
the Bell-Kochen-Specker contradiction were built involving Sign
and Reflection observables\cite{fleming} and also for unitary
operators, functions of position and momentum\cite{clifton}. In
all these cases, the operators involved are not trivial functions
of position and momentum and it would be convenient to devise a
simple proof. In this work, the incompatibility of quantum
mechanics with the assumption of the existence of noncontextual
values for position and momentum of just one particle is shown.
The proof is very simple (\emph{a posteriori}) and it involves
the simplest physical system (one spinless free particle) and
therefore it makes the Bell-Kochen-Specker result easier
accessible to non experts.

In this work we will investigate the possibility of existence of
definite values for position and momentum observables of a single
particle. Of course these putative values are not provided by
quantum mechanics and it does not matter whether they are
deterministic, as in a hidden variable theory, or are random
values distributed according to some inherent randomness in
nature (zitterbewegung). In an ensemble of systems, we only
require that the proposed putative values should be distributed
according to the distributions predicted by quantum mechanics.
\section{THE PUTATIVE VALUE}
Let us assume that we can assign to any Hilbert space operator $A$
a numerical value $\overline{A}$ called the \emph{putative value},
with the following properties.
\begin{itemize}
    \item \emph{Completeness}: the set $\{\overline{A}\}$ of all
    possible putative values is the spectrum of the corresponding
    operator.
    \item  \emph{Functional consistency:} the putative value
    preserves functional relations in the sense that for any
    function $F$  it is  $\overline{F(A)}=F(\overline{A})$.
    \item  \emph{Context independence:} the putative value
    assumed by an operator is independent of the context in
    which the corresponding observable is placed. Different contexts
    are defined by different sets of commuting operators.
\end{itemize}
It can be proved\cite{ish} that the functional consistency
condition has the important consequence that the putative value
for \emph{commuting} operators are additive and multiplicative.
That is,
\begin{equation}\label{aditandmul}
   [A,B]=0\  \rightarrow\ \overline{A+B}=
   \overline{A}+\overline{B}\ \mbox{ and }\
   \overline{A\cdot B}=\overline{A}\cdot \overline{B}\ .
\end{equation}
These relations can be generalized to functions that can be
expanded as power series: $\overline{
F(A,B)}=F(\overline{A},\overline{B})$, however we will not use
the general form. In fact we will only need the additive
property. This additive property is not necessarily true for non
commuting operators. For instance if $A=J_{x}$ and $B=J_{y}$ are
two components of angular momentum, then $A+B=\sqrt{2}J_{u}$ is
also a component of angular momentum in a different direction but
the spectrum of $\sqrt{2}J_{u}$ is clearly not equal to the sum
of the spectra of $J_{x}$ and $J_{y}$.

We will now investigate whether it is possible to assign putative
values $\overline{X}$ and $\overline{P}$ to the position and
momentum observables $X$ and $P$ of a particle, in a way
compatible with quantum mechanics. This compatibility means that
the putative values should be distributed according to the
probability functions provided by quantum mechanics. Of course,
quantum mechanics can not predict or compute these putative
values but we want to know if their \emph{existence} is allowed
by quantum mechanics. We just want to see if we can \emph{think}
that these values exist. It is also irrelevant whether these
values can be calculated by a deterministic hidden variables
theory or they are assigned randomly.

Let us assume that the position observable is divided by some
length scale $\lambda$ (Compton length, for instance) and
momentum is multiplied by $\lambda/\hbar$ making them
dimensionless. Therefore their associated values are pure numbers
and the addition of position with momentum is not meaningless.
Let us consider the operators $X_{1},X_{2},P_{1},P_{2}$
corresponding to the observables of position and momentum of a
particle in a plane and let
$\overline{X}_{1},\overline{X}_{2},\overline{P}_{1},\overline{P}_{2}$
be their putative values. Let us now build several linear
combinations of these operators that can be grouped in
intersecting subsets of commuting operators. In Fig.1 we see some
of these linear combinations and we notice that all operators
joined by a straight line commute. Let us consider the two
operators $A=X_{1}-X_{2}+P_{1}+P_{2}$ and
$B=X_{1}+X_{2}+P_{1}-P_{2}$. Since they commute, their putative
values are such that
\begin{equation}\label{putval1}
 \overline{A+B}=\overline{A}+\overline{B}\ .
\end{equation}
The left hand side of this equation is
$\overline{A+B}=\overline{2X_{1}+2P_{1}}$, and the right hand
side is
$\overline{A}+\overline{B}=\overline{X_{1}-X_{2}+P_{1}+P_{2}}
+\overline{X_{1}+X_{2}+P_{1}-P_{2}}
=\overline{X_{1}-X_{2}}+\overline{P_{1}+P_{2}}
+\overline{X_{1}+X_{2}}+\overline{P_{1}-P_{2}} =
\overline{X}_{1}-\overline{X}_{2}+\overline{P}_{1}+\overline{P}_{2}
+\overline{X}_{1}+\overline{X}_{2}+\overline{P}_{1}-\overline{P}_{2}
=2\overline{X}_{1}+2\overline{P}_{1}$. Therefore the equation
above becomes
\begin{equation}\label{putval2}
\overline{X_{1}+P_{1}}=\overline{X}_{1}+\overline{P}_{1} \ .
\end{equation}
We have used the additive property of the putative values of
\emph{commuting} observables and we have shown that, even though
$X_{1}$ and $P_{1}$ \emph{do not commute}, their putative values
are also additive. This is indeed very suspicious and for most
experts this would be sufficient reason to deny the existence of
the putative values. Anyway we will prove that this result is in
contradiction with quantum mechanics, but before doing this, some
comments are convenient.
\begin{figure*}
\setlength{\unitlength}{1cm}
\begin{picture}(20,8)
\thicklines \put(4,1){\framebox(9,4)} \put(4,3){\line(1,0){9}}
\multiput(4,1)(3,0){4}{\circle*{0.3}}
\multiput(4,5)(3,0){4}{\circle*{0.3}}
\multiput(4,3)(9,0){2}{\circle*{0.3}}
\put(3.3,5.5){$\mathbf{X_{1}-X_{2}}$}
\put(6.8,5.5){$\mathbf{X_{1}}$} \put(9.8,5.5){$\mathbf{X_{2}}$}
\put(12.3,5.5){$\mathbf{X_{1}+X_{2}}$}
\put(3.3,0.3){$\mathbf{P_{1}+P_{2}}$}
\put(6.8,0.3){$\mathbf{P_{1}}$} \put(9.8,0.3){$\mathbf{P_{2}}$}
\put(12.3,0.3){$\mathbf{P_{1}-P_{2}}$}
\put(-0.3,3){$\mathbf{A=X_{1}-X_{2}+P_{1}+P_{2}}$}
\put(13.5,3){$\mathbf{B=X_{1}+X_{2}+P_{1}-P_{2}}$}
\end{picture}
\caption[FIGURE 1.]{ Set of operators used to show that, although
X and P do not commute, their putative values are additive.
Notice that all operators joined by a straight line
commute.\vspace{2cm}}
\end{figure*}
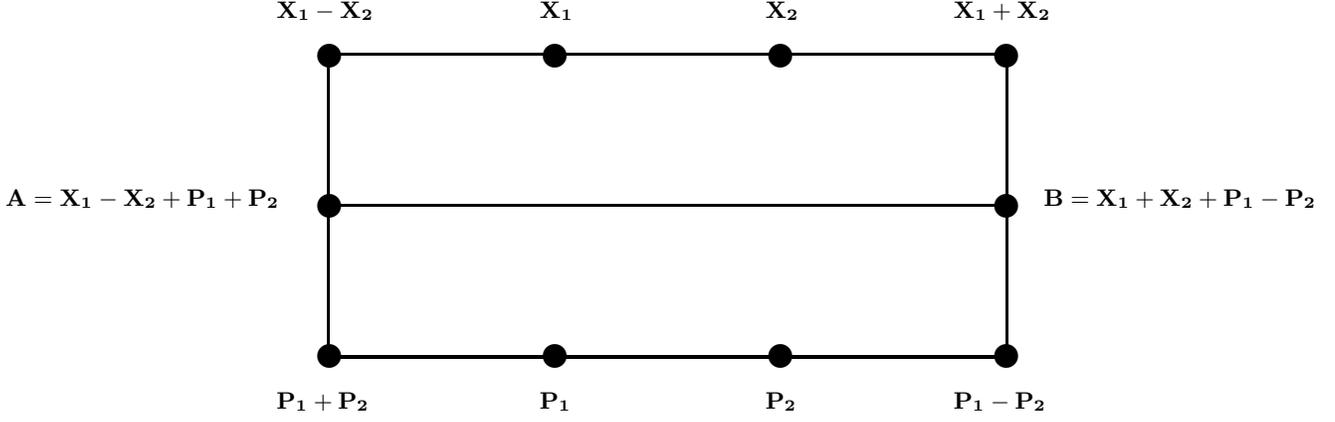

The property of context independence is necessary in the above
argument because we assume that the value of, for instance, the
operator $X_{1}+X_{2}$ in the upper right corner of Fig.1, is the
same when we consider it as a member of the set
$\{X_{1}-X_{2},X_{1},X_{2},X_{1}+X_{2}\}$ as the value it takes
when it is a member of the set $\{X_{1}+X_{2},B,P_{1}-P_{2}\}$.
Without this assumption, Eq.(\ref{putval2}) could not be obtained.
We could have taken other sets of commuting operators leading to
similar results. For instance in the top line of Fig.1 we could
take the operators $X_{1}+P_{2},X_{1},P_{2},X_{1}-P_{2}$ and the
appropriate set in the lower line. Also, choosing the signs
properly, instead of an addition we could get a substraction in
Eq.(\ref{putval2}) or any linear combination of the operators.
Considering the projection of position and momentum along two
arbitrary orthogonal directions in three dimensional space, would
lead us to the conclusion that for any linear combination we have
$\overline{\alpha\textbf{X}+\beta\textbf{P}}=
\alpha\overline{\textbf{X}}+\beta\overline{\textbf{P}}$. Notice
that in order to obtain these results it is important that the
commutator of $X$ and $P$ is a \emph{constant} and a subtle
cancellation of the commutator in different directions is made.
This cancellation is no longer possible when we use a similar
scheme in order to try to prove that
$\overline{X^{2}+P^{2}}=\overline{X^{2}}+\overline{P^{2}}$.
Unfortunately several attempts to prove this failed. If we could
prove this, then we would obtain an immediate contradiction
because the spectrum of $X^{2}+P^{2}$ is discrete whereas the
spectra of $X^{2}$ and $P^{2}$ are continuous. An immediate
contradiction would also follow if we could prove that
$\overline{XP}=\overline{X}\ \overline{P}$ because in one case the
spectrum is complex and in the other it is real.
\section{CONTRADICTION WITH QUANTUM MECHANICS}
If the putative values for position $X$ and momentum $P$ exist,
then they must be such that, for the operator $S=X+P$, we have
$\overline{S}=\overline{X+P}=\overline{X}+\overline{P}$. We will
now see that this is in contradiction with quantum mechanics. For
this, let us consider an ensemble of systems described by quantum
mechanics by a Hilbert space element $\psi$. Let
$\{\varphi_{x}\}\ ,\{\phi_{p}\}\ ,\{\eta_{s}\}$ be the
eigenvectors of the operators $X,P,S$ corresponding to the
eigenvalues $x,p,s$. According to quantum mechanics, these three
observables will have the probability distribution functions
\begin{eqnarray}
  \rho(x) &=& |\langle\varphi_{x},\psi\rangle|^{2} \ ,\\
  \varpi(p) &=& |\langle\phi_{p},\psi\rangle|^{2}\ , \\
  \sigma(s)&=& |\langle\eta_{s},\psi\rangle|^{2}\ .
\end{eqnarray}
The assumptions made are the usual ones when we deal with
position and momentum observables. However in order to be more
rigourous we should state that the operators $X,P$ and $S$ are
unbound (this follows from their commutation relations) and they
have no eigenvectors \emph{in} the Hilbert space. The solution to
this problem is to define a \emph{Rigged} Hilbert space (known as
a Gel'fand triplet in mathematics) that contains the sets of
generalized eigenvectors $\{\varphi_{x}\}\ ,\{\phi_{p}\}\
,\{\eta_{s}\}$ that can be used as basses in order to expand any
Hilbert space element\cite{bal}. The modulus squared of the
expansion coefficients are interpreted in quantum mechanics as
the probability distributions given in the equations above. As an
example, the generalized eigenvectors in the coordinate
representation, where $X=x$ and $P=-i\partial_{x}$, are given by
\begin{eqnarray}
  \varphi_{x_{0}}(x) &=& \delta (x-x_{0}) \ ,\\
   \phi_{p}(x) &=& \frac{1}{\sqrt{2\pi}}\exp\left(ipx\right)\ , \\
 \eta_{s} (x)&=& \frac{i^{1/4}}{\sqrt{2\pi}}
 \exp\left(-i\left(\frac{(x-s)^{2}}{2}-\frac{s^{2}}{4}\right)\right)\
 .
\end{eqnarray}
One can easily check that they satisfy their corresponding
eigenvalue equations and that they are ``delta function''
normalized. We will not use these functions but we present them
just in order to clarify that the probability distributions in
Eqs.(4) to (6) are mathematically well defined.

In the ensemble of systems, the putative values of position and
momentum $\overline{X}$ and $\overline{P}$ are distributed
according to $\rho(x)$ and $\varpi(p)$, then the addition of
these two random variables
$\overline{S}=\overline{X+P}=\overline{X}+\overline{P}$ is
distributed, according to the theory of random variables, by the
convolution
\begin{eqnarray}
  \sigma_{pv}(s)&=& \int\!\!dx\ \rho(x)\ \varpi(s-x) \nonumber \\
    &=& \int\!\!\!dx\!\!\!\int\!\!\!dp\ \rho(x)\ \varpi(p)
    \ \delta\left(s-(x+p)\right)\ .
\end{eqnarray}
Now we will see that this putative value prediction for the
distribution is different from the quantum mechanical prediction
in Eq.(6), that can be written as
\begin{equation}\label{sigmamq}
   \sigma(s)= \int\!\!\!dx\!\!\!\int\!\!\!dp\
\langle\varphi_{x},\psi\rangle \langle\psi,\phi_{p}\rangle\
\langle\phi_{p},\eta_{s}\rangle\langle\eta_{s},\varphi_{x}\rangle
   \ ,
\end{equation}
and appears formally quite different from the putative value
distribution. The formal difference between these two
distributions is such that, presumably,
$\sigma_{pv}(s)\neq\sigma(s)$ for all states $\psi$. In
particular, it is easy to show that for some states both
distributions have different dispersion, that is,
$\Delta^{2}_{pv}(s)\neq\Delta^{2}_{mq}(s)$. The quantum
mechanical prediction is:
\begin{eqnarray*}
  \Delta^{2}_{mq}(s) &=& \langle S^{2}\rangle-\langle S\rangle^{2} =
  \langle (X+P)^{2}\rangle-\langle X+P\rangle^{2}\\
    &=& \langle X^{2}+P^{2}+XP+PX\rangle-\langle X\rangle^{2}-\langle P\rangle^{2}
    -2\langle X\rangle\langle P\rangle\\
   &=& \Delta^{2}(x) + \Delta^{2}(p)+
   \langle XP+PX\rangle-2\langle X\rangle\langle P\rangle\ ,
\end{eqnarray*}
and the putative value prediction is
\begin{eqnarray*}
  \Delta^{2}_{pv}(s) &=&\int\!\!\!ds\  s^{2} \sigma_{pv}(s)
  -\left(\int\!\!\!ds\  s\ \sigma_{pv}(s)\right)^{2}\\
&=&\int\!\!ds\  s^{2} \int\!\!\!dx\!\!\!\int\!\!\!dp\ \rho(x)\
\varpi(p)\ \delta\left(s-(x+p)\right)
  -\left(\int\!\!\!ds\  s  \int\!\!\!dx\!\!\!\int\!\!\!dp\  \rho(x)\ \varpi(p)
    \ \delta\left(s-(x+p)\right) \right)^{2} \\
&=&  \int\!\!\!dx\!\!\!\int\!\!\!dp\  (x+p)^{2}\rho(x)\ \varpi(p)
      -\left(\int\!\!\!dx\!\!\!\int\!\!\!dp\  (x+p)\ \rho(x)\ \varpi(p)
    \right)^{2} \\
&=&  \int\!\!\!dx\!\!\!\int\!\!\!dp\ (x^{2}+p^{2}+2xp)\ \rho(x)\
\varpi(p)
  -\left(\int\!\!\!dx\ x\ \rho(x)+ \int\!\!\!dp\ p\ \varpi(p)
  \right)^{2}\\
   &=& \Delta^{2}(x) + \Delta^{2}(p)\ .
\end{eqnarray*}
Clearly, at least for any state with non vanishing
correlation\cite{rob} $\langle XP+PX\rangle-2\langle
X\rangle\langle P\rangle\neq 0$, the assumption of the existence
of the putative values is in contradiction with quantum
mechanics.
\section{CONCLUSION}
If the predictions of quantum mechanics are correct, then we have
proved that position and momentum observables can not be assigned
any context independent value. The proof presented here involves
elementary observables of one single particle and provides a very
simple illustration of the Bell-Kochen-Specker contradiction.

Context \emph{dependent} putative values are not prohibited and
all attempts to replace standard quantum mechanics by some form
of hidden variables theories must necessarily include the context
dependence in the deterministic assignment of values to the
observables. This necessity makes such deterministic theories
less appealing. One of the main reasons for developing hidden
variables theories was to bring the quantum world closer to the
classical expectations but the necessary contextuality goes in
the other direction.

\section{Acknowledgements}
I would like to thank M. Hoyuelos and A. Jacobo for comments.
This work received partial support from ``Consejo Nacional de
Investigaciones Cient{\'\i}ficas y T{\'e}cnicas'' (CONICET), Argentina.

\end{document}